\documentclass[12pt]{iopart}
\usepackage{epsfig}  
\usepackage{wrapfig}
\begin{document}

\title[First results from NA60]{First results from NA60 on low
  mass muon pair production in In-In collisions at 158~GeV/nucleon}

\author{S.~Damjanovic$^{1}$ (for the NA60 Collaboration)}
\vspace*{-0.5cm}
\author{
  R~Arnaldi$^{2}$, R~Averbeck$^{3}$, K~Banicz$^{1}$, K~Borer$^{4}$, 
  J~Buytaert$^{5}$, J~Castor$^{6}$, B~Chaurand$^{7}$, W~Chen$^{8}$, 
  B~Cheynis$^{9}$, C~Cicalo$^{10}$,  A~Colla$^{2}$, P~Cortese$^{2}$,
  A~David$^{5,11}$, A~De~Falco$^{10}$, N~de Marco$^{2}$, A~Devaux$^{6}$, 
  A~Drees$^{3}$, L~Ducroux$^{9}$, H~En'yo$^{12}$, A~Ferretti$^{2}$, 
  M~Floris$^{10}$, A~F\"orster$^{5}$, P~Force$^{6}$, A~Grigorian$^{13}$, 
  J-Y~Grossiord$^{9}$,  N~Guettet$^{5,6}$, A~Guichard$^{9}$, 
  H~Gulkanian$^{13}$, J~Heuser$^{12}$, M~Keil$^{1,5}$, L~Kluberg$^{7}$, 
  Z~Li$^{8}$, C~Louren\c{c}o$^5$, J~Lozano$^{11}$, F~Manso$^{6}$, 
  P~Martins$^{5,11}$,  A~Masoni$^{10}$, A~Neves$^{11}$, H~Ohnishi$^{12}$, 
  C~Oppedisano$^{2}$, P~Parracho$^{11}$, P~Pillot$^{9}$, G~Puddu$^{10}$, 
  E~Radermacher$^{5}$, P~Ramalhete$^{11}$, P~Rosinsky$^{5}$, 
  E~Scomparin$^{2}$, J~Seixas$^{11}$, S~Serci$^{10}$, R~Shahoyan$^{5,11}$, 
  P~Sonderegger$^{11}$, H~J~Specht$^{1}$,  R~Tieulent$^{9}$, E~Tveiten$^{5}$, 
  G~Usai$^{10}$, H~Vardanyan$^{13}$, R~Veenhof$^{11}$ and H~W\"ohri$^{5,11}$
}

\vspace*{0.5cm}

\address{
$^{1}$ Universit\"{a}t Heidelberg, Heidelberg, Germany

$^{2}$ Universit\`a di Torino and INFN, Italy

$^{3}$ SUNY Stony Brook, New York, USA

$^{4}$ Laboratory for High Energy Physics, University of Bern, Bern, Switzerland

$^{5}$ CERN, Geneva, Switzerland

$^{6}$ LPC, Universit\'e Blaise Pascal and CNRS-IN2P3, Clermont-Ferrand, France

$^{7}$ LLR, Ecole Polytechnique and CNRS-IN2P3, Palaiseau, France

$^{8}$ Brookhaven National Laboratory, Upton, New York, USA

$^{9}$ IPN-Lyon, Universit\'e Claude Bernard Lyon-I and CNRS-IN2P3, Lyon, France

$^{10}$ Universit\`a di Cagliari and INFN, Cagliari, Italy

$^{11}$ Instituto Superior T\'ecnico, Lisbon, Portugal

$^{12}$ RIKEN, Wako, Saitama, Japan

$^{13}$  YerPhI, Yerevan Physics Institute, Yerevan, Armenia
}

\begin{abstract}
The NA60 experiment at the CERN SPS studies dimuon production in
proton-nucleus and nucleus-nucleus collisions. The combined
information from a novel vertex telescope made of radiation-tolerant silicon
pixel detectors and from the muon spectrometer previously used in NA50
allows for a precise measurement of the muon vertex and a much
improved dimuon mass resolution. We report on first results from the
data taken for Indium-Indium collisions at 158 AGeV/nucleon in 2003,
concentrating on a subsample of about 370\,000 muon pairs in the mass
range $<1.2$~GeV/$c^{2}$. The light vector mesons $\omega$ and $\phi$
are completely resolved, with a mass resolution of about
23~MeV/$c^{2}$ at the $\phi$. The transverse momentum spectra of the
$\phi$ are measured over the continuous range $0<p_{\rm T}<2.5$~GeV/c;
the inverse slope parameter of the spectra is found to increase with
centrality, with an average value of $T=252\pm3$~MeV.
\end{abstract}




\section{Introduction}
The NA60 experiment was approved in the year 2000 to study a number of
physics topics accessible through the measurement of muon pairs. All
topics were studied by previous experiments, but were left with major
open questions: (i)~the excess emission of lepton pairs for masses
below the $\rho$-resonance with the possible link to the chiral transition,
(ii)~the enhanced production of intermediate mass muon pairs with the
ambiguity of either prompt thermal radiation or increased open charm
production, and (iii)~the precise mechanism underlying J/$\psi$ suppression, 
asking for a variation of the system size. This paper is restricted to 
the first topic, while the second and third are treated in another
NA60 paper presented during this conference~\cite{ho:el}.
  
The NA45/CERES experiment has consistently observed an excess emission
of electron pairs with invariant masses $0.2<m<0.6$~GeV/$c^{2}$ above
known sources from hadron decays, in S-Au and Pb-Au collisions, which
is concentrated at low pair $p_{\rm T}$ and scales steeper than linear
with the associated charged particle multiplicity~\cite{ceres:el,
ceres1:el}. The theoretical interpretation of this excess has been
linked to the restoration of chiral symmetry in the hot and dense
medium, leading to a ``melting'' of the $\rho$ with extra yield at lower masses
and decreased yield at the nominal pole
position~\cite{theo:el}.  However, statistical accuracy and mass
resolution up to and including the 2000 data have not been sufficient to
positively verify this scenario; the excess continuum seems
structureless up to the $\rho$/$\omega$ region, and even
$q\overline{q}$ annihilation cannot be ruled out at present. Better
statistics, signal-to-background ratio and mass resolution are
therefore required to clarify the existing ambiguities. The NA60
experiment has now potentially achieved this goal. However, only
preliminary results can be presented at this stage, including raw
spectra over the whole mass region. More detailed results are only
given on the properties of the $\phi$.

\section{Experimental set-up}
The essential components of the NA60 experiment are shown in figure~1.
\begin{figure}
\begin{center}
\mbox{\epsfig{file=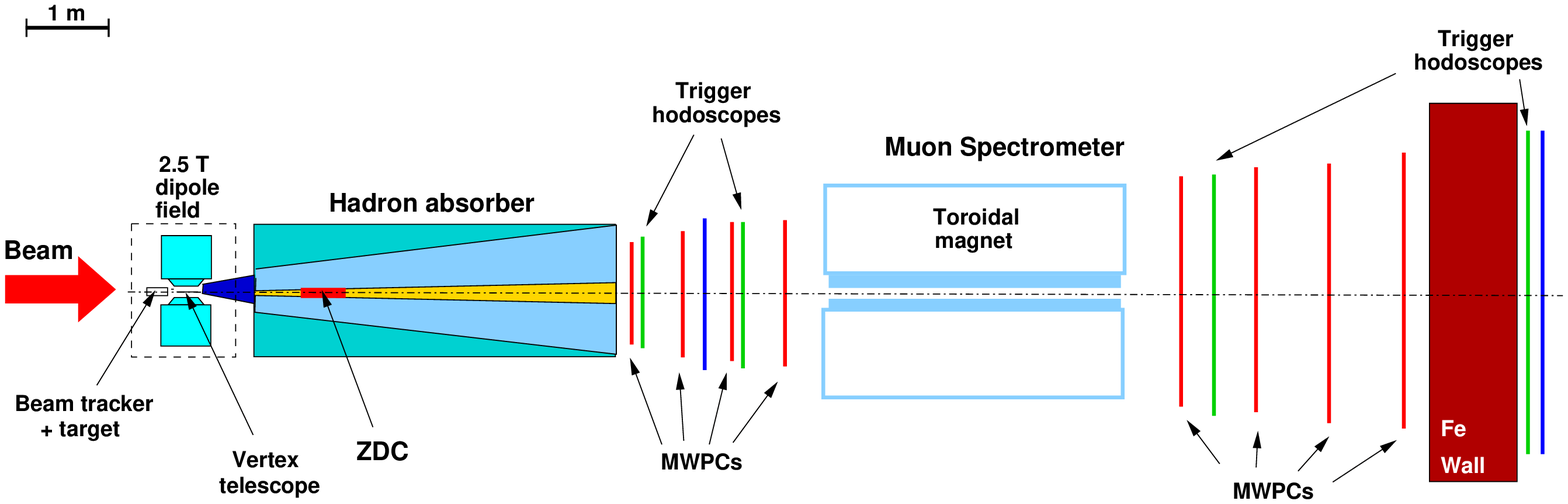,width=0.8\textwidth}}
\mbox{\epsfig{file=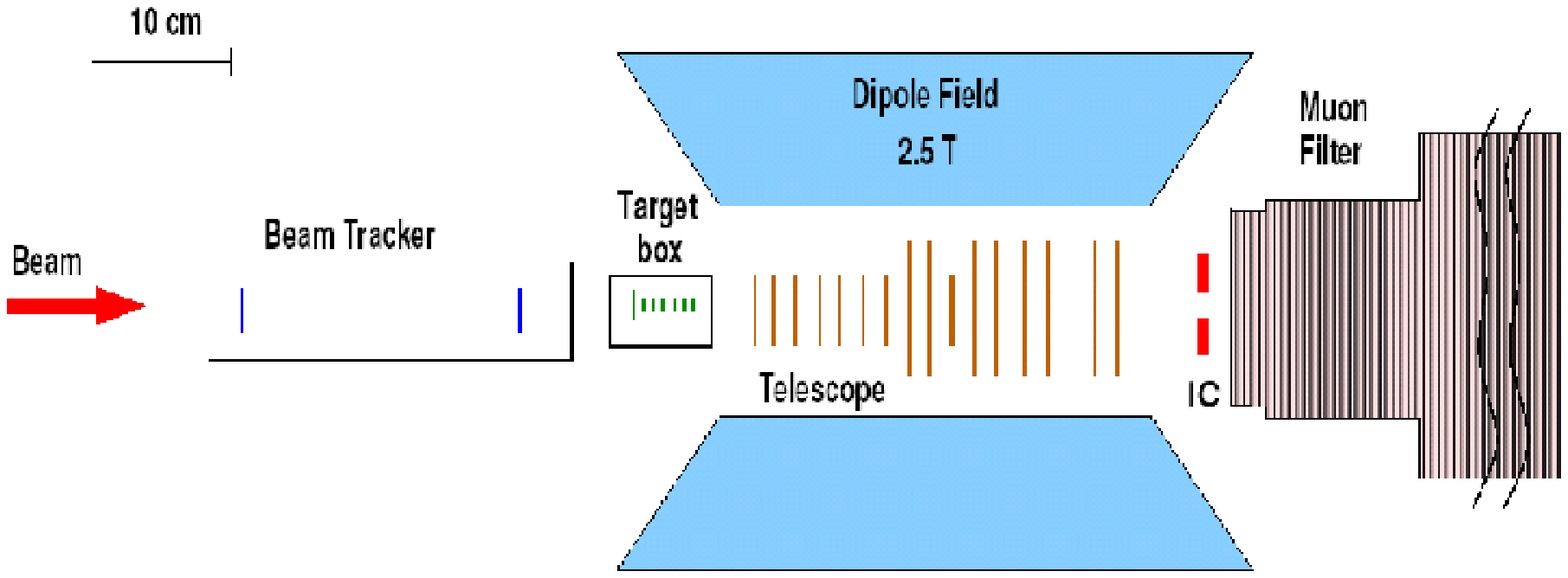,width=0.6\textwidth}}
\caption{Full NA60 set-up (upper) and detail of the target region (lower).}
\label{fig:setup}
\end{center}
\end{figure} 
The muon spectrometer previously used in NA10/NA38/NA50
consists of an air-core toroidal magnet, 8 multi-wire proportional
chambers for muon tracking and momentum determination, and 4
scintillator hodoscopes for the muon pair trigger. A 5.5~m long hadron
absorber before the spectrometer serves as the muon filter, with
the usual drawbacks of such a set-up: energy loss and multiple
scattering of the muons impair the mass resolution and prohibit an
accurate vertex determination. The new silicon pixel telescope added
by NA60 is used to track all charged particles in the vertex region
before the hadron absorber and to determine their momenta independently
of the muon spectrometer. It consists of a number of tracking planes
with a total of 12 space points, embedded in a 2.5~T dipole
magnet. The planes are made from assemblies of detector chips bump-bonded
to radiation-tolerant readout chips, developed for the ALICE and LHCb
experiments at the LHC~\cite{pixel:el}. Each chip contains
$256\times32$ pixels with a pixel size of
$50\times425$~$\mu$m$^{2}$. Matching of the muon tracks before and
after the hadron absorber both in {\it coordinate and momentum} space
improves the mass resolution in the region of the light vector mesons
$\omega$, $\phi$ by a factor of nearly 3, decreases the combinatorial
background by (partial) kink-rejection of muons from $\pi$ and
K~decays, and finally allows the identification of displaced vertices of muons
from D-decays for the measurement of open
charm~\cite{ho:el,ruben:el}.

Further components of the NA60 set-up are a beam tracker of 4
cryogenic silicon microstrip detectors upstream of the target to track
the incoming beam particles, and a zero-degree quartz-fiber ``spaghetti''
calorimeter (``ZDC'') to tag the centrality (number of participants)
of the collision. The high luminosity of NA50 is kept in NA60. Radiation
tolerance and high readout speed of the silicon pixel telescope allow
for beam intensities of $5\cdot10^{7}$ per 5~s~burst for ions
and $2\cdot10^{9}$ per~5~s~burst for protons in connection with target
thicknesses of more than 10\,\% of an interaction lenght. The dimuon
trigger is effective enough to record the resulting data rates without
any centrality selection.

\section{Global results for low mass muon pairs}

During the 5-week long Indium run in 2003, around $4\cdot10^{12}$
ions with an energy of 158~GeV/nucleon were delivered at the target,
and a sample of 230~Million dimuon events (background dominated)
were written to tape. The presently available dimuon mass spectra,
without any centrality selection, are displayed in figure~2.
\begin{figure}
\mbox{\epsfig{file=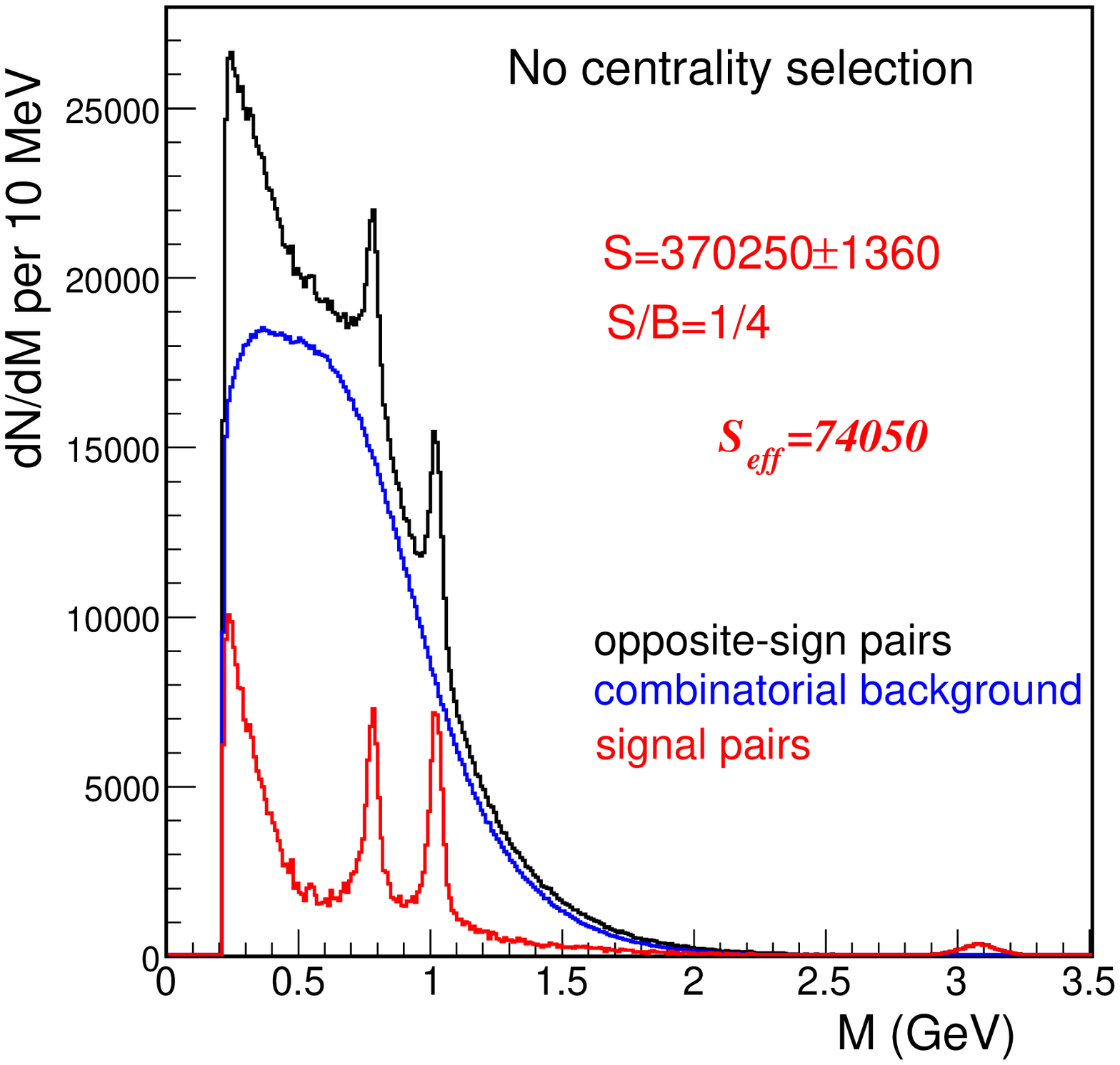,width=0.52\textwidth}}
\mbox{\epsfig{file=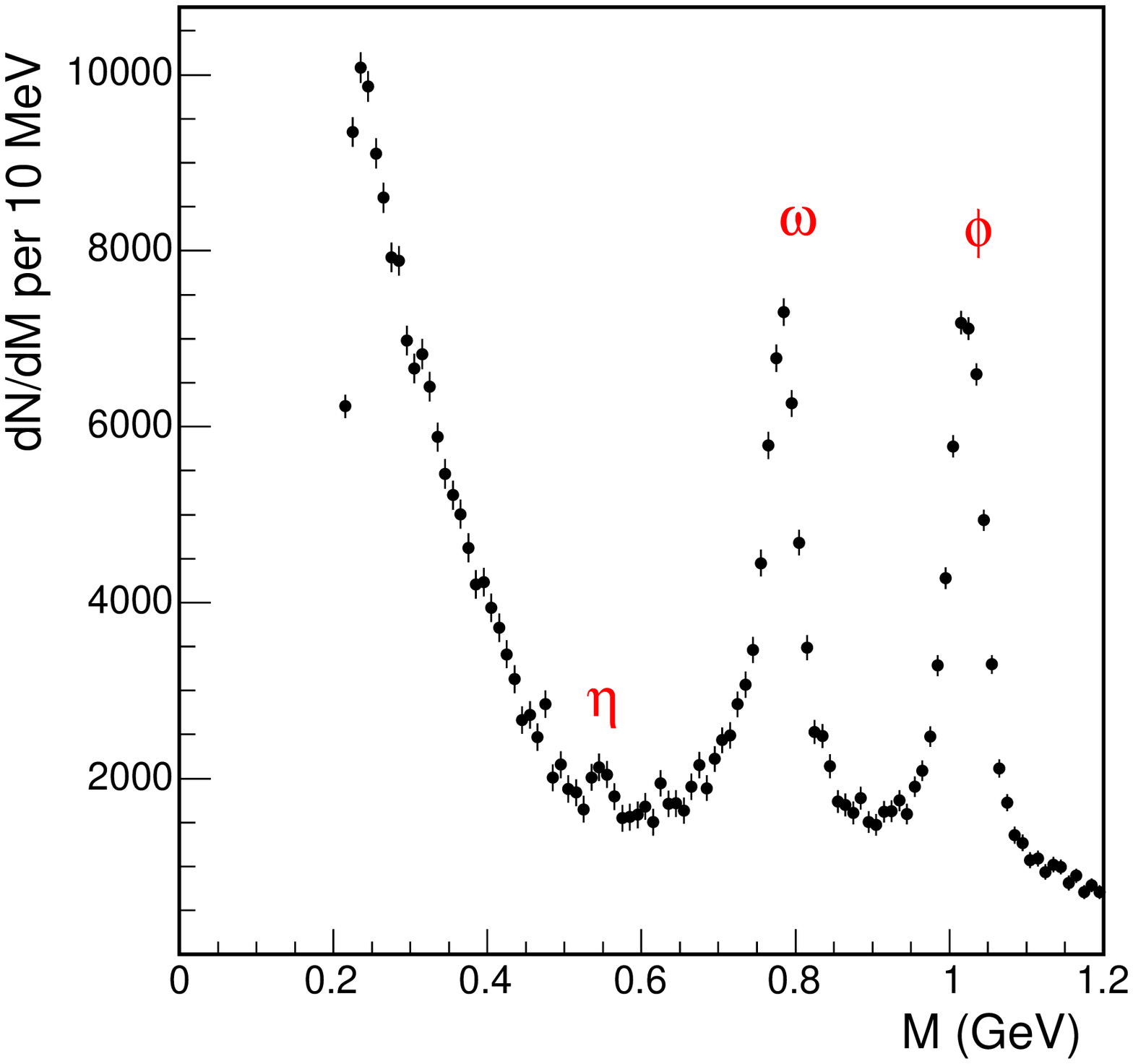,width=0.55\textwidth}}
\caption{Invariant mass spectra for measured opposite-sign muon pairs,
  combinatorial background and net pairs after background subtraction
  (left), and enlarged section of the net pairs (right).}
\label{fig:mass}
\end{figure} 
The combinatorial background of opposite-sign muon pairs is determined
by mixing single muons from events with like-sign pairs; subtraction
of this background from the measured opposite-sign pairs results in
the net spectrum labeled ``signal pairs''. The mean
signal-to-background ratio is about 1/4. The net spectrum contains
370\,000 pairs, corresponding to about 35\% of the available total
statistics; the final sample will therefore contain about 1 Million
pairs with an effective statistics of 10$^{6}$/(4+1) =
200\,000. Compared to the CERES Pb-Pb data 1995/96~\cite{ceres:el} or
2000~\cite{ceres1:el}, this is an improvement by a factor of roughly
1\,000. The enlarged low mass net spectrum on the right of figure~2
shows the light vector mesons $\omega$ and $\phi$ to be completely
resolved. The mass resolution for the $\phi$ is about 23 MeV/c$^{2}$,
independent of centrality. One can even recognize the rare
$\eta\rightarrow\mu\mu$ decay, which should lead to the first
unambiguous cross section measurement of the $\eta$ in nuclear collisions. 

It should be stressed that the extension of the mass spectrum all the
way down to the 2m$_{\mu}$ threshold is accompanied by a complete
coverage in pair transverse momentum down to zero, albeit with
decreasing acceptance by up to 2 orders of magnitude for the lowest
masses. This presents a further drastic improvement compared to the
NA50 set-up.

\begin{figure}[h!]
\begin{center}
\mbox{\epsfig{file=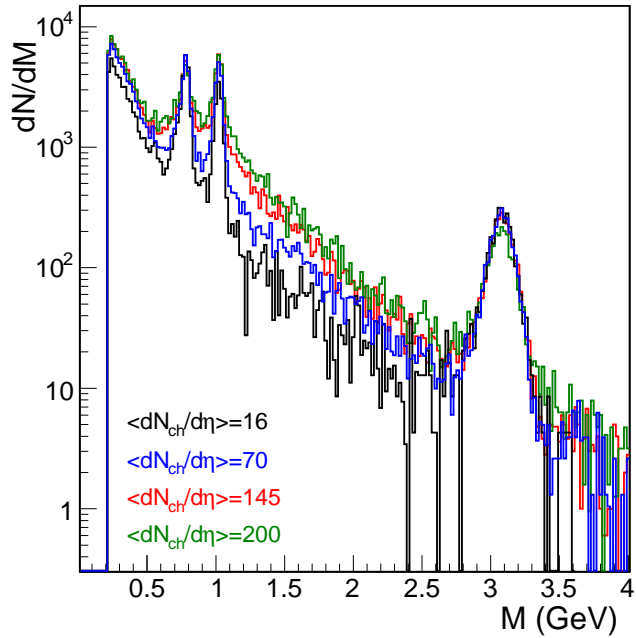,width=0.6\textwidth}}
\caption{Invariant mass spectra for 4 bins in associated charged particle multiplicity.}
\label{fig:masslog}
\end{center}
\end{figure}
Although the statistics is high enough for finer binning, the data
have so far only been subdivided into 4 coarse bins in associated
charged particle multiplicity, as measured by the pixel telescope. The
net invariant mass spectra for these 4 bins are shown in figure~3,
arbitrarily normalized in the region of the $\omega$ peak. 
One recognizes some relative variation at very low masses and about a
factor of 2 increase in the $\phi$. Stronger variations occur below
the $\omega$, between the $\omega$ and the $\phi$, and above the
$\phi$. Those significantly above the $\phi$ are mostly due to the
fact that hard processes scale with the number of binary
nucleon-nucleon collisions
rather than with $N_{\rm part}$ (like the $\omega$). The quantitative
judgment of the rest has to await a remaining correction to the mass
spectra of figures~2 and~3 which has not yet been done: the subtraction
of muon pairs arising from incorrect (``fake'') matches of muon tracks
between the two spectrometers. This correction will also be obtained
from mixing events, but the corresponding software has not yet been
finalized. From overlay MC simulations it seems, however, that the
level of fake matches is too small to account for the varying yield of
dimuons in the neighborhood of the $\omega$ and $\phi$~\cite{ruben:el}.

\section{Detailed results for the $\phi$ meson}

With the $\omega$ and $\phi$ so well resolved and isolated from the
rest of the low mass pairs, a more detailed analysis has already been
performed on the $\phi$/$\omega$ yield ratio and the
transverse momentum distribution of the $\phi$ as a function of
centrality. The analysis is based on the sample shown in figure~2, 
containing about 37\,000 net events for the $\phi$ (14\,000 effective
statistics), after subtraction of the remaining underlying continuum.
\begin{figure}
\begin{center}
\mbox{\epsfig{file=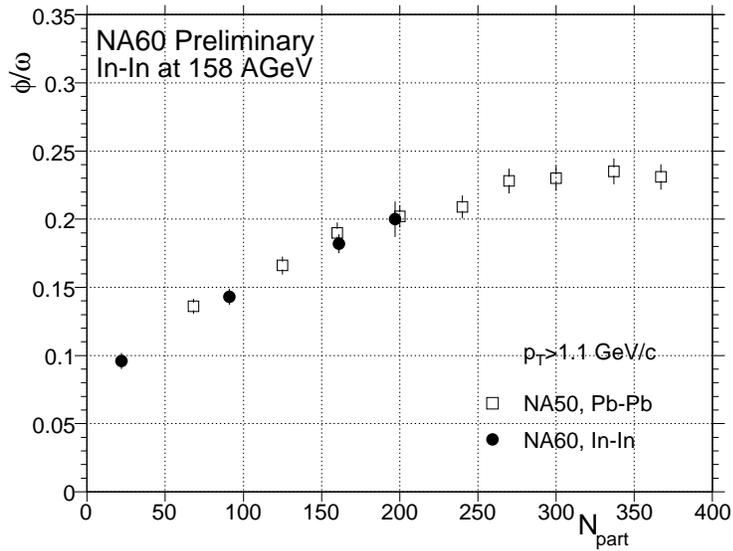,angle=0,width=0.65\textwidth}}
\caption{Ratio of the cross sections for $\phi$ and $\omega$
  production in the rapidity window $3.3<y<4.2$. The NA60 data are
  obtained from the procedure described in the text. The quoted errors 
  are purely statistical. The data from NA50 shown for comparison
  are also discussed in the text.}
\label{fig:yield}
\end{center}
\end{figure}
The $\phi/\omega$ yield ratio has been determined from the raw data by
propagating the muon pairs arising from the known
resonance~($\eta,\rho,\omega,\phi$) and
Dalitz~($\eta,\eta^{\prime},\omega$) decays through the NA60 set-up
with GEANT, using the hadron-decay generator GENESIS~\cite{gen:el} as
an input. An additional continuum source with
exponential fall-off beyond the $\phi$ and the level of fake matches
as obtained from overlay MC simulations have been taken into account
in an approximate way. The final result on $\phi/\omega$ is obtained
by adjusting the input ratio $\phi/\omega$ of GENESIS such that the
output fits the measured data. It should be stressed that the results
are rather insensitive to the details of the procedure, with
one exception: the $\rho/\omega$ ratio in the fit procedure is
somewhat ill-defined, while the sum $\rho+\omega$ is stable.  For this
reason, the fit has been constrained to a fixed ratio
$\rho/\omega=$~1. The whole procedure is done independently for each of
the 4 multiplicity bins. The average multiplicity density in each of
the bins is converted, via the correlation of multiplicity vs.\
ZDC~energy, to average values for the number of participants $N_{\rm
part}$, using Glauber fits to the ZDC data. The results for
$\phi/\omega$ are plotted in figure~4. Note that the errors are purely
statistical; the systematic errors are under investigation. Results
previously obtained by NA50 for the system Pb-Pb are shown for
comparison. They have been derived from the published values of
$\phi/(\rho+\omega)_{\mu\mu}$~\cite{na50:el}, again assuming 
$\rho/\omega=$~1 to be consistent, and correcting for the branching
ratios of $\rho,\omega,\phi\rightarrow\mu\mu$. They are also
corrected to correspond to the same cut $p_{\rm T}>1.1$~GeV/$c$ for
the $\rho,\omega,\phi$ rather than the common cut $m_{\rm T}>1.5$~GeV/$c^{2}$~\cite{na50:el}, 
using the NA50 slope parameter value $T=228$~MeV for the extrapolation. 
It is remarkable that the two data sets agree within errors, both in
absolute value and in the slope~vs.~$N_{\rm part}$. This implies
$N_{\rm part}$ to be a reasonable scaling variable for particle ratios
in different collision systems, at least as long as A is not too small.

\begin{figure}
\begin{center}
\mbox{\epsfig{file=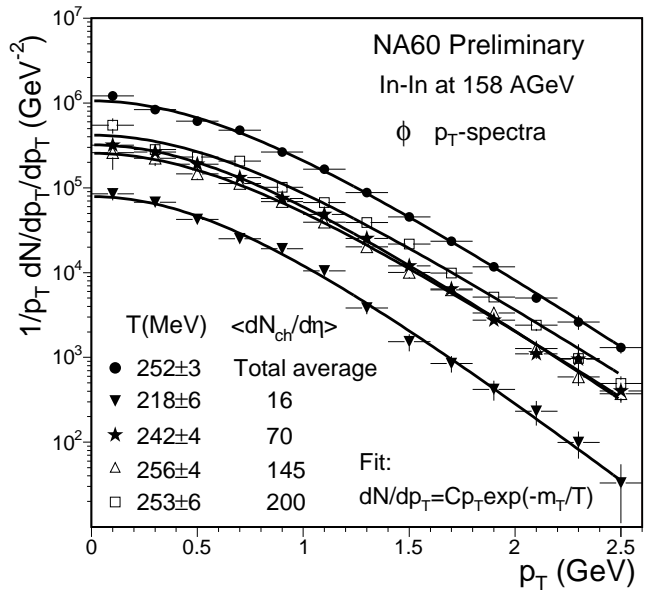,width=0.6\textwidth}}
\caption{Transverse momentum spectra of the $\phi$ meson for
  4 bins in associated charged multiplicity density. The errors are purely
  statistical; the systematic errors are under investigation.}
\label{fig:pt}
\end{center}
\end{figure}
The analysis of the transverse momentum spectra of the $\phi$ has been
done in a straightforward way. The raw data are obtained by selecting
a narrow window around the $\phi$ peak in the net spectrum of figure~2,
and then subtracting the content of 2 side-windows symmetrically placed
around the peak. It was verified that the results are completely
insensitive to the choice of width and position of the
side-windows, up to the extreme of applying the whole procedure to the
gross spectrum before subtraction of the combinatorial background. The
raw data were then corrected for acceptance, using a 2-dimensional
acceptance matrix in $p_{\rm T}$ and $y$ obtained from a detailed
simulation of the NA60 set-up, and finally converted to invariant
relative cross sections $1/p_{\rm T} {\rm d}N/{\rm d}p_{\rm T}$. The
acceptance correction is quite uncritical, since the acceptance varies
by less than a factor of 2 over the whole $p_{\rm T}$ range. Finally,
the invariant cross sections were fitted with the simple form
$\exp(-m_{\rm T}/T)$ to extract the slope parameters $T$ in a way
consistent with other publications. The spectra and associated
$T$ values show no variation with $y$ within errors. The
$y$-integrated data for $3.3<y<4.1$ are plotted in figure~5 separately
for the 4 bins
in associated charged particle multiplicity density. The data extend
up to $p_{\rm T}=2.5$~GeV/$c$; for the full-statistics sample, a limit of
3.5~GeV/$c$ may be reachable. The slope parameters derived from
the fits are well defined within the chosen parametrization. The
average slope parameter over the whole range in multiplicity and
$p_{\rm T}$ is $T=252\pm3$~MeV; if the fit is restricted to $p_{\rm
T}<1.5$~GeV/$c$ (NA49 range) or $m_{\rm T}>1.5$~GeV/$c^{2}$ (NA50
range), the resulting values are $256\pm3$ and $245\pm5$, respectively, i.e.\
nearly identical within errors.

The tendency of the slope parameter $T$ to rise with $\langle {\rm
d}N_{\rm ch}/{\rm d}\eta \rangle$ or $N_{\rm part}$ is clearly borne
out in figure~6.
\begin{figure}
\begin{center}
\mbox{\epsfig{file=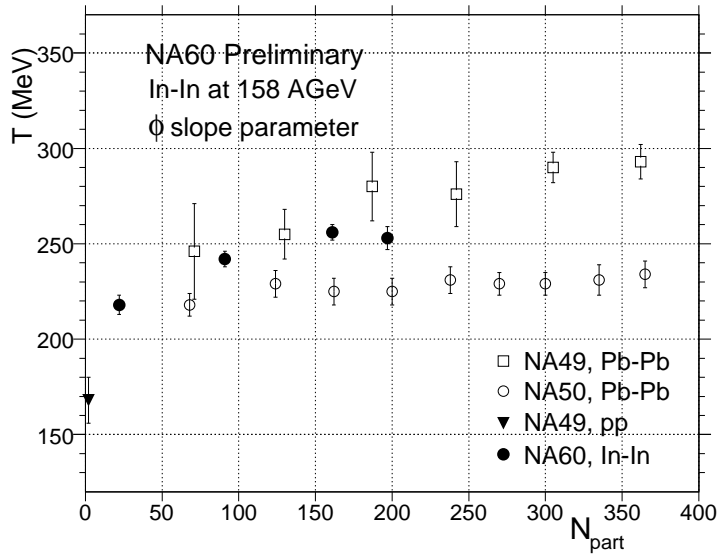,width=0.65\textwidth}}
\caption{Slope parameters $T$ of the transverse momentum spectra of
  the $\phi$ meson for different experiments at 158~GeV/nucleon. The
  errors shown for NA60 are purely statistical; the systematic
  errors are under investigation, but are presently believed to be
  $\leq10$~MeV.}
\label{fig:pt}
\end{center}
\end{figure}
For comparison, this plot also contains the slope parameters reported
by NA49 for Pb-Pb on $\phi\rightarrow$ KK~\cite{na49:el}, and those
from NA50 for Pb-Pb on
$\phi\rightarrow\mu\mu$~\cite{na50:el}. Remarkably, NA60
and NA49 agree in the region of overlap in $N_{\rm part}$
within the rather large errors of NA49. Whether that bears on the
famous ``$\phi$-puzzle'', originally discussed in view of the
discrepancy between NA49 and NA50 for Pb-Pb~\cite{shuryak:el}, remains
to be seen. A difference now also exists between NA60 and NA50 for
which we have no obvious explanation, and the usefulness of $N_{\rm
part}$ as the proper scaling variable between different systems for
quantities other than particle ratios is in any case not proven. A
consistent solution of the $\phi$-puzzle by NA60 would require
parallel data on $\phi\rightarrow$ KK for the In-In system. Such an
analysis, based solely on track information from the pixel telescope,
is indeed in progress. Work on a precision determination of the mass
and width of the $\phi$, which addresses further aspects of in-medium
effects on the $\phi$, is also in progress.

\section{Conclusions}

The NA60 experiment is setting new standards in the data quality of
muon pair measurements. A high statistics run with protons on many
different nuclear targets is presently being performed to provide
precision reference data for all three major aspects of the NA60
program. Specifically, the low mass region will benefit from 
unprecedented sensitivity to sources other than the known meson
decays.

\section*{Acknowledgments}

The speaker and the Heidelberg group wish to thank
the German BMBF for financial support. 

\end{document}